\documentclass[aps,twocolumn,prd,nofootinbib,superscriptaddress,preprintnumbers,10pt]{revtex4-2}

\usepackage[utf8]{inputenc}

\usepackage{mathtools}
\usepackage{amsfonts}
\usepackage{amssymb}
\usepackage{mathrsfs}
\usepackage{bbm}
\usepackage{braket}
\usepackage{slashed}
\usepackage{tensor}

\usepackage{graphicx}
\usepackage{array}

\usepackage{placeins}
\usepackage{makecell}
\usepackage[caption=false]{subfig}
\usepackage{float}

\usepackage{xspace}
\usepackage{xfrac}
\usepackage{hyperref}
\usepackage[nameinlink]{cleveref}
\usepackage{appendix}
\usepackage{units}

\usepackage{xifthen}
\usepackage{booktabs}
\usepackage{multirow}
\usepackage{courier}
\usepackage[dvipsnames]{xcolor}
\hypersetup{
	colorlinks,
	linkcolor={blue!75!black},
	citecolor={blue!75!black},
	urlcolor={blue!75!black}
}

\Crefname{equation}{Eq.\!}{Eqs.\!}

\DeclareMathOperator{\Tr}{Tr}
\newcommand{\insertedOp}[0]{\mathcal{Q}}

\graphicspath{{./figs/}}

\newcommand{\gettitle}{Variance reduction in lattice QCD observables via normalizing flows}

\newcommand{\getMITAffiliation}{\affiliation{Center for Theoretical Physics, Massachusetts Institute of Technology, Cambridge, MA 02139, USA}}
\newcommand{\getIAIFIAffiliation}{\affiliation{The NSF AI Institute for Artificial Intelligence and Fundamental Interactions}}
\newcommand{\getFNALAffiliation}{\affiliation{Fermi National Accelerator Laboratory, Batavia, IL 60510, U.S.A.}}
\newcommand{\getEdinburgAffiliation}{\affiliation{Higgs Centre for Theoretical Physics, School of Physics and Astronomy,\\
University of Edinburgh, EH9 3FD Edinburgh, United Kingdom}}
\newcommand{\getBernAffiliation}{\affiliation{Albert Einstein Center, Institute for Theoretical Physics, University of Bern, 3012 Bern, Switzerland}}

\hypersetup{
    pdftitle={\gettitle},
    pdfauthor={Abbott,Boyda,Fu,Hackett,Kanwar,Romero-López,Shanahan,Urban},
    pdfkeywords={lattice field theory}{lattice gauge theory}{normalizing flows}{machine learning},
    bookmarksopen=true,
    bookmarksopenlevel=2,
    bookmarksnumbered=true
}

\begin{document}

\title{\gettitle}

\author{Ryan~Abbott}
\affiliation{Physics Department, Columbia University, New York, NY 10027, USA}
\getMITAffiliation
\getIAIFIAffiliation
\author{Denis~Boyda}
\getMITAffiliation
\getIAIFIAffiliation
\author{Yang Fu}
\getMITAffiliation
\getIAIFIAffiliation
\author{Daniel~C.~Hackett}
\getFNALAffiliation
\author{Gurtej~Kanwar}
\getEdinburgAffiliation
\author{Fernando~Romero-L\'opez}
\getBernAffiliation
\author{Phiala~E.~Shanahan}
\getMITAffiliation
\getIAIFIAffiliation
\author{Julian~M.~Urban}
\getMITAffiliation
\getIAIFIAffiliation

\preprint{FERMILAB-PUB-26-0130-T, MIT-CTP/6010}

\begin{abstract}
    Normalizing flows can be used to construct unbiased, reduced-variance estimators for lattice field theory observables that are defined by a derivative with respect to action parameters. This work implements the approach for observables involving gluonic operator insertions in the SU(3) Yang-Mills theory and two-flavor Quantum Chromodynamics (QCD) in four space-time dimensions.
    Variance reduction by factors of $10$--$60$ is achieved in glueball correlation functions and in gluonic matrix elements related to hadron structure, with demonstrated computational advantages. The observed variance reduction is found to be approximately independent of the lattice volume, so that volume transfer can be utilized to minimize training costs.
\end{abstract}

\maketitle

\section{Introduction}\label{sec:intro}

Lattice field theory provides a powerful numerical framework for computing physical observables in Quantum Chromodynamics (QCD), the fundamental theory of the strong interaction. Lattice QCD studies span from precision determinations of fundamental parameters of the Standard Model, to properties of nuclear systems. The approach is, however, notoriously computationally demanding: calculations require both the generation of gauge-field configurations via sophisticated sampling algorithms and the subsequent evaluation of observables through expensive measurement procedures. Improvements in either the sampling algorithms or the measurement methods can directly increase the scope and precision of lattice QCD~\cite{Joo:2019byq}.

Machine-learned normalizing flows have been proposed as tools to address both challenges~\cite{Albergo:2019eim}. From the perspective of sampling, using flows for gauge-field generation, 
replacing or supplementing more conventional Hybrid Monte Carlo approaches, has been shown to alleviate critical slowing down and improve topological sampling~\cite{Albergo:2022qfi,Abbott:2024mix,Bonanno:2025pdp}. From the perspective of measurement, recent work~\cite{Bacchio:2023all,Catumba:2025ljd,Abbott:2024kfc} has proposed using flows to reduce the variance of observables that can be written as derivatives with respect to action parameters, with the first demonstration in QCD reported in Ref.~\cite{Abbott:2025kvi}. This methodology utilizes flows to compute numerical derivatives by bridging between nearby distributions. While progress on both fronts remains rapid, applications to date have been demonstrated at scales and/or in regions of parameter space considerably different from the state-of-the-art regime of lattice QCD. Specifically, there have not yet been results demonstrating significant computational advantage for the setting of greatest phenomenological interest: calculations using large lattice volumes, for a variety of observables, in QCD with dynamical fermions.

This work demonstrates an application of flows that achieves significant variance reduction in several phenomenologically relevant applications in both dynamical QCD and Yang–Mills (YM) theories in four space-time dimensions. The study implements residual-flow architectures, introduced in Refs.~\cite{Abbott:2023thq,Abbott:2024kfc}, and is deployed for the first time for lattice geometries up to $16^3 \times 32$. Two categories of observables are investigated: (1) glueball correlation functions and (2) gluonic matrix elements in the context of hadron structure; in both cases the relevant observables can be expressed as derivatives with respect to action parameters. Variance reduction by a factor of $10$--$60$ is observed across all cases. The implications of this reduction for the cost and workflow of lattice QCD calculations is discussed in detail. Importantly, the variance reduction does not degrade with volume transfer, which enables efficient training by volume-transfer approaches.

Addressing the bias of finite-difference approximations in derivative observables, this work also introduces a method to linearize learned flow transformations, yielding instantaneous flow fields that can be used to exactly evaluate the derivative with respect to action parameters. It is demonstrated that this unbiased estimate is equivalent to the control variate method of Refs.~\cite{Bacchio:2023all,Catumba:2025ljd}, in this case using the efficient masked residual-flow architecture and optimizing using the reverse Kullback-Leibler divergence as an objective function.
It is proven and demonstrated numerically that, at leading order in the finite-difference parameter, the unbiased linearized estimator is identical to the finite-difference estimator on each gauge configuration. These estimates therefore reproduce the same variance reduction when flows are trained for short-distance transformations. At leading order, training short-distance residual flows with the reverse Kullback-Leibler (KL) divergence is equivalent to training flow fields directly using a Physics-Informed Neural Network (PINN) loss function~\cite{Albergo:2024trn}, suggesting a future workflow in which both training and evaluation are obtained in this exact limit.

This work is organized as follows. First, the notion of derivative observables and how flows can be applied to reduce their variance is introduced in \Cref{sec:corrflows}; \Cref{sec:workflow} summarizes the proposed workflow for such calculations.
\Cref{sec:glueballs} presents applications to glueball correlation functions, while \Cref{sec:structure} demonstrates applications for hadron structure. 
In \Cref{sec:advan}, the potential computational advantage of the methodology is discussed. 
\Cref{sec:concl} concludes and presents an outlook on future applications. The connection between the KL divergence and PINN loss is detailed in \Cref{app:training}.

\section{Flows for variance reduction}\label{sec:corrflows}

This section describes the notion of derivative observables and how flow models can be used to reduce their variance.

\subsection{Derivative observables}

Many observables of interest in lattice QCD can be obtained by studying the derivative of path-integral quantities with respect to the action parameters. Two distinct classes of such derivative observables are considered in this work.

\paragraph{Vacuum-subtracted correlation functions via the derivative trick.}
A broad class of $N$-point correlation functions can be constructed from $(N-1)$-point functions by inserting an operator ${\insertedOp}$ into the action and applying the `derivative trick' as follows. Assuming an action of the form $S_\lambda = S_0 - \lambda {\insertedOp}$, where $S_0$ is the unperturbed action (in this case that of Yang-Mills theory or QCD), expectation values at a given value of $\lambda$ are defined as
\begin{equation}
\begin{aligned}
   \langle \mathcal{O} \rangle_\lambda &\equiv \frac{1}{\mathcal{Z}_\lambda} \int DU\, \mathcal{O}(U)\, e^{-S_\lambda(U)}, \\
   \quad \mathcal{Z}_\lambda &\equiv \int \mathcal{D}U \, e^{-S_\lambda(U)}.
\end{aligned}
\end{equation}
In the following, $\langle \cdot \rangle_0$ will be used to denote expectation values in the unperturbed ($\lambda = 0$) ensemble.
When $\mathcal{O}$ is a product of $N-1$ localized operators, taking a derivative of $\left< \mathcal{O} \right>$ with respect to $\lambda$ yields a vacuum-subtracted $N$-point correlation function:
\begin{equation}
    \langle \mathcal{O}\, {\insertedOp} \rangle_0 - \langle \mathcal{O} \rangle_0 \langle {\insertedOp} \rangle_0\,=\frac{d \langle \mathcal{O} \rangle_\lambda}{d\lambda} \bigg\rvert_{\lambda=0} .\label{eq:derivative_trick}
\end{equation}
This is a specific application of the generating functional approach in quantum field theory, which for example also provides a route to computing susceptibilities and response functions from derivatives of expectation values.

\paragraph{Hadronic matrix elements via the Feynman--Hellmann theorem.}
A second application of the same underlying idea allows the extraction of hadronic matrix elements (see e.g., Ref.~\cite{QCDSF:2012mkm}). Again taking $S_\lambda = S_0 - \lambda {\insertedOp}$, the Feynman--Hellmann theorem relates the matrix element of ${\insertedOp}$ in a hadronic state $|h\rangle$ at $\lambda = 0$ to the $\lambda$-dependence of the hadron energy $E_h(\lambda)$, extracted from two-point correlation functions computed with the modified action:
\begin{equation}
    \langle h |\, {\insertedOp}\, | h \rangle -\langle \Omega |\, {\insertedOp}\, | \Omega \rangle = 2\, E_h\, \frac{\partial E_h(\lambda)}{\partial \lambda} \bigg\rvert_{\lambda=0}\,, \label{eq:FH}
\end{equation}
where $|\Omega \rangle$ is the interacting vacuum state.
Unlike the derivative trick, which involves differentiating the correlator directly, the Feynman--Hellmann approach requires first extracting the energy eigenvalue from the spectral decomposition of the two-point function, then differentiating with respect to $\lambda$.
This difference in the order of operations---differentiating the extracted energy rather than the correlator itself---cleanly isolates the matrix element from changes in the overlap factors, and is a practical distinction between the two approaches.

\subsection{Variance of derivative observables}

The derivative trick provides an interesting perspective into the variance of derivative observables~\cite{Catumba:2023ulz,Catumba:2025ljd}, or equivalently, why $N$-point functions are more noisy than $(N-1)$-point functions. This can be seen by writing out the derivative trick as an explicit limit of a finite difference,
\begin{equation}
    \frac{d \langle \mathcal O \rangle_\lambda }{d\lambda} \Bigg \rvert_{\lambda=0} = \lim_{\lambda \to 0} \frac{\langle  \mathcal O \rangle_\lambda - \langle  \mathcal O \rangle_0}{\lambda}. \label{eq:fdiff0}
\end{equation}
The standard approach to estimating the $N$-point function is equivalent to evaluating $\left< \mathcal{O} \right>_\lambda$ in the above expression by reweighting from expectation values under the $\lambda=0$ action:
\begin{equation}
\begin{aligned}
    \left< \mathcal{O} {\insertedOp} \right>_0 - \left<\mathcal{O}\right>_0 \left<{\insertedOp}\right>_0
    &= \lim_{\lambda \to 0}
    \left< \mathcal{O} \, \left( \frac{w^{\rm Id}_\lambda - 1}{\lambda}\right) \right>_0 \\
    &= \left< \mathcal{O} \, \partial_\lambda w^{\rm Id}_\lambda \right>_0 ,
\end{aligned}
\label{eq:fdiff1}
\end{equation}
where the normalized reweighting factor is
\begin{align}
    w^{\rm Id}_\lambda(U) &= \frac{\mathcal{Z}_0}{\mathcal{Z}_{\lambda}} \hat{w}^{{\rm Id}}_\lambda(U) = \frac{\hat{w}^{\rm Id}_\lambda(U)}{\left< \hat{w}^{\rm Id}_\lambda \right>_0}
\end{align}
in terms of the unnormalized weights
\begin{equation}
    \hat{w}^{{\rm Id}}_\lambda(U) = e^{\lambda {\insertedOp}(U)}.
\end{equation}
Here the superscript ``Id'' denotes that this is a reweighting factor directly from $S_0$ to $S_\lambda$, without first applying a flow transformation to the field sampled from $S_0$, as is considered in future sections.
The unsubtracted correlation function can be obtained by using the unnormalized weighting factors instead, as
\begin{equation}
\begin{aligned}
    \langle \mathcal{O} {\insertedOp} \rangle_0 &= \lim_{\lambda \to 0}
    \left< \mathcal{O} \, \left( \frac{\hat{w}^{\rm Id}_\lambda - 1}{\lambda} \right) \right>_0 \\
    &= \left< \mathcal{O} \, \partial_\lambda \hat{w}^{\rm Id}_\lambda \right>_0 . \label{eq:fdiff2}
\end{aligned}
\end{equation}
\Cref{eq:fdiff1,eq:fdiff2} can be confirmed by evaluating
\begin{equation}
\begin{aligned}
    \frac{\partial}{\partial \lambda} w^{\rm Id}_\lambda \Big|_{\lambda = 0} &= {\insertedOp} - \langle {\insertedOp} \rangle_0, \quad
    \frac{\partial}{\partial \lambda} \hat{w}^{\rm Id}_\lambda \Big|_{\lambda = 0} = {\insertedOp}.
\end{aligned}
\end{equation}

From the expression in \Cref{eq:fdiff1}, one sees that the subtracted $N$-point function obtained via the derivative trick can be expressed as the $(N-1)$-point observable reweighted by $\partial_\lambda w^{\rm Id}_\lambda$. Under the assumption that $\mathcal O$ and $\partial_\lambda w^{\rm Id}_\lambda$ are approximately independent\footnote{Whether this is a good approximation depends strongly on the observable, and its validity cannot be established in general terms.}, the variance of the derivative observable can be approximated as:
\begin{equation}
    {\rm Var}\left[ \frac{d \langle \mathcal O \rangle }{d\lambda} \right] \simeq  {\rm Var}\left[ \partial_\lambda w^{\rm Id}_\lambda \right]  \langle \mathcal O^2 \rangle  + \cdots,
    \label{eq:vardO}
\end{equation}
where the ellipsis indicates corrections due to correlations between the observable, its square and the reweighting factors. 
In the finite-difference approximation, the relevant variance is
\begin{equation} \label{eq:E2}
    \mathcal{E}^2_{{\rm Id}} := {\rm Var} \left[ \frac{w^{\rm Id} - 1}{\lambda} \right],
\end{equation}
for which $\lim_{\lambda \to 0} \mathcal{E}^2_{{\rm Id}} = {\rm Var}[\partial_\lambda w^{\rm Id}_\lambda]$,
so that the variance of the derivative operator is proportional to $\mathcal{E}^2_{{\rm Id}}$ under the independence assumption. 
This conclusion provides an obvious path forward: modifying the reweighted estimate, for example by incorporating flow transformations, to reduce fluctuations of the reweighting factor around 1 should lead to a lower variance in the observable.

\subsection{Flows}

A flow~\cite{rezende2016variational, dinh2017density, JMLR:v22:19-1028} is defined as a diffeomorphism $f$ that maps samples from a base distribution $r(U)$ to a model distribution with density
\begin{equation} \label{eq:flow-density}
    q(V) = r(U) \times  J(V) \ ,  \quad V=f(U),
\end{equation}
where the Jacobian of the flow transformation is defined as
\begin{equation}
    J(V) \equiv \left|\det \frac{\partial f(U)}{\partial U}\right|^{-1}.
\end{equation}
The flow $f$ is typically constructed with many free, trainable parameters, which are optimized such that the obtained model distribution approximates a target distribution $p$, i.e., $q(V)\simeq p(V)$.

In the context of sampling lattice QCD gauge fields~\cite{Abbott:2023thq}, the samples $U$ and $V$ are lattice gauge-field configurations, and the probability distributions $p$ and $r$ are defined in terms of Euclidean lattice actions: 
\begin{equation} \label{eq:prior-target-densities}
    r(U) \propto e^{-S(U)}, \quad p(V) \propto e^{-S'(V)}.
\end{equation}
Here, $S$ and $S'$ represent the fact that the Euclidean actions have different parameters, and the proportionality is up to the normalization constant of the distribution. Any flow $f$ can be used to construct unbiased estimates according to the target distribution by reweighting samples according to the weights $w(V) = p(V)/q(V)$, which can be explicitly computed using \Cref{eq:flow-density,eq:prior-target-densities}.

There are several options to optimize the parameters defining the flow, all of which aim to minimize the difference between the model density $q(V)$ and the target density $p(V)$. For example, one can perform stochastic gradient descent on a direct estimate of the Kullback-Leibler (KL) divergence~\cite{Kullback:1951}, using reduced-noise estimates of the gradients with path gradients~\cite{vaitl2022gradients} or related methods~\cite{Abbott:2023thq}, or avoiding gradients of $p(V)$ using the ``REINFORCE'' algorithm~\cite{Bialas:2023fyj}. This work uses the path gradient approach to minimize the reverse KL divergence.

A common construction of expressive flows is based on the composition of invertible layers,
\begin{equation}
    f = g_1 \circ g_2 \circ ... \circ g_n \ .
\end{equation}
Several gauge-equivariant architectures for the invertible layers $g_k$ were discussed in Ref.~\cite{Abbott:2023thq}. This work uses a modified version of the residual layers described in Ref.~\cite{Abbott:2024kfc}, which are based on the trivializing map~\cite{Luscher:2009eq}.  Similar architectures were also used in Refs.~\cite{Nagai:2021bhh,Bacchio:2022vje,Bacchio:2023all,Gerdes:2024rjk,Nagai:2025rok}.

Each layer transforms a subset of gauge links, defined as ``active", conditioned on the remaining ``frozen" links. The choice of partitioning is referred to as the masking pattern, which can be specified as $x_\mu \in \{0,1\}$, where 0 and 1 respectively denote frozen and active links at space-time position $x$ in direction $\mu$. In this work, ``mod $N$'' masking patterns are used, which satisfy the condition  
${(p+\sum_\mu x_\mu) = 0\; (\mathrm{mod} \, N)}$ for parities $p \in \{0,..,N-1\}$. A  ``mod 2'' masking pattern, for instance, corresponds to a checkerboard.

Active links in residual layer $k$ are transformed as
\begin{equation}\label{eq:reslayer}
    U'_\mu(x) = e^{\lambda F^{k}_x(U_f, U_\mu(x))} U_\mu(x) \ ,
\end{equation}
where the ``flow field'' of the layer $F^k_x(U_f, U_\mu(x))$ is formed from linear combinations of algebra-projected, untraced Wilson loops that begin and end at the position $x$. This construction ensures gauge equivariance, since under a gauge transformation with local parameter $\Omega_x$, the field transforms as $F^k_x \to \Omega_x F^k_x \Omega_x^\dagger$.
Here an explicit factor of $\lambda$, which defines the target,  is included based on the empirical observation that inductive bias results in residual flow transformations that tend to scale as $O(\lambda)$, for small $\lambda$. This fact is used to efficiently linearize these flows in \Cref{sec:lin} below.

The coefficients used to build the flow field out of untraced Wilson loops are learnable. Schematically:
\begin{equation}
    \lambda F^k_x(U_f, U_\mu(x)) = \sum_n \alpha_n \mathcal{P}[U_\mu(x) S^{n}_{x,\mu\nu}(U_f)], \label{eq:force}
\end{equation}
where $\mathcal{P}[W]$ is a projection to anti-Hermitian, traceless elements of the algebra, $S^n_{x,\mu\nu}$ is a staple-shaped object that depends only on the frozen links, and $\alpha_n$ are trainable coefficients. In addition, a convolution is applied to the frozen links to increase the expressivity---see  Ref.~\cite{Abbott:2024kfc} for the precise implementation and Ref.~\cite{Favoni:2020reg} for a related architecture.

In the architecture presented in Ref.~\cite{Abbott:2024kfc}, translation invariance is used in the coefficients in the flow field, i.e., $\alpha_n$ in \Cref{eq:force} are not position dependent. This increases training stability and reduces memory costs. However, some of the target distributions in this work have less symmetry, e.g.\ if a timeslice operator is inserted as in \Cref{eq:interpglue}. To effectively learn such targets, a dependence on the timeslice is introduced as $\alpha_n(x_0)$, explicitly breaking time-translation invariance in the parameters defining the flow field. This only leads to a moderate increase of training costs and memory usage.

It is also necessary to be able to evaluate the Jacobian determinant defining $J(V)$. The masking pattern applied to construct the transformation in \Cref{eq:reslayer} ensures that the Jacobian matrix is triangular with respect to the indices $x$. For the $k$-th residual layer, only links in the $\mu$ direction are changed. The determinant of the Jacobian matrix of this layer is then entirely determined by the diagonal blocks and can be evaluated as
\begin{equation}
    J_{k}(V) = \prod_{x \, \text{active}} \left|\det_{ab} \nabla^a_{x,\mu} \Tr[-2 T^b (\lambda F^k_x)] \right|^{-1}.
\end{equation}
Here the symbol $\nabla$ denotes the left Lie derivative,
$\nabla_{x,\mu}^a\,\phi(U) = \frac{d}{dt}\phi(\ldots, e^{tT^a}U_\mu(x),\ldots)|_{t=0}$ and the generators are taken to be anti-Hermitian and normalized according to $\Tr[T^a T^b] = -\tfrac{1}{2} \delta^{ab}$.
The full Jacobian is obtained by composing the Jacobian determinant of all layers.
In this work, autodifferentiation is applied to evaluate the Jacobian factor for each active link.

\subsection{Derivative observables via flows} \label{subsec:derivviaflow}
A natural way to use flows in the computation of derivative operators is to consider the case where $S=S_0$ and $S'=S_\lambda =S_0 -\lambda {\insertedOp}$. Finite-difference approximations of the derivatives appearing in the right-hand sides of \Cref{eq:derivative_trick} and \Cref{eq:FH} may then be calculated by using flows to evaluate unbiased expectation values $\left< \cdot \right>_\lambda$ and the corresponding unflowed fields to evaluate $\left< \cdot \right>_0$.
For the derivative trick, this leads to
\begin{align}
\frac{d\langle\mathcal O \rangle}{d\lambda}\Bigg|_{\lambda=0} &\simeq
\frac{1}{\lambda} \Big\langle w_\lambda(V) \, \mathcal{O}(V) - \mathcal{O}(U) \Big\rangle_0,
\label{eq:dOdLflow}
\end{align}
where $V = f(U)$ is distributed according to $q(V)$, and the reweighting factor is defined as
\begin{equation} \label{eq:rw-lambda}
    w_\lambda(V) = \frac{\hat{w}_\lambda(V)}{\left< \hat{w}_\lambda(V) \right>_q}, \quad \hat{w}_\lambda(V) = e^{-S_\lambda(V)} / q(V).
\end{equation}
In the Feynman--Hellmann application, the right-hand side of \Cref{eq:FH} may be similarly evaluated as
\begin{equation}
    \frac{\partial E_h}{\partial \lambda} \bigg \rvert_{\lambda = 0} \simeq \frac{1}{\lambda}\big( E_h(\lambda) - E_h(0) \big). \label{eq:dMpi0}
\end{equation}
The perturbed energy $E_h(\lambda)$ is determined by fitting the standard spectral decomposition to the two-point hadronic correlation function $C^h_\lambda (t)$ computed on the flowed fields: 
\begin{align}
\begin{split}
    C^h_\lambda (t) &=\Big \langle O^h_t(U) \, [O^h_0(U)]^\dagger   \Big \rangle_\lambda \\ &= \Big \langle w_\lambda(V) \, O^h_t(V) \, [O^h_0(V)]^\dagger  \Big \rangle_0, \label{eq:Ch}
\end{split}
\end{align}
where $O^h_t$ and $[O^h_0]^\dagger$ respectively indicate an annihilation operator at Euclidean time $t$ and a creation operator at Euclidean time $0$ with quantum numbers of the state $|h\rangle$. Evaluating this correlation function on the flowed fields implies evaluating all gluonic quantities and all quark propagators using the flowed gauge configurations $V$.
\Cref{sec:lin} discusses an alternative to the finite-difference approximation which has particular advantages when computing correlation functions with multiple choices of the inserted operator. 

The overall quality of a given flow model is often quantified with the Effective Sample Size per configuration (ESS). In this context, it is defined as
\begin{equation}\label{eq:ESSgeneral}
    \text{ESS } 
    = \frac{\braket{w_\lambda}_q^2}{\braket{w_\lambda^2}_q}
    = \frac{1}{1 + {\rm Var}[w_\lambda]}.
\end{equation}
The ESS can be estimated using $N$ gauge field configurations generated from $q(V)$,
\begin{equation}
    {\rm ESS} \approx \frac{1}{N} \frac{\left[ \sum_{i=1}^N w_\lambda(V_i)  \right]^2}{\sum_{i=1}^N \big [w_\lambda(V_i)\big]^2} \ ,
\end{equation}
with the value of this estimator lying in the interval ${\text{ESS} \in [\tfrac{1}{N}\, ,\, 1]}$. A perfect model corresponds to ${{\rm Var}[w_\lambda] = 0}$ and ${\text{ESS}=1}$.

In these applications, where a flow model is used to map between arbitrarily similar distributions (i.e., for infinitesimally small $\lambda$), the ESS can be arbitrarily close to 1, depending on the choice of $\lambda$ itself. A more useful metric for model quality is then $\mathcal{E}^2$, as defined in \Cref{eq:E2} but for the trained flow, which can be related to the ESS according to
\begin{equation}
    \mathcal{E}^2 = \frac{1}{\lambda^2} \left( \frac{1}{\rm ESS} - 1 \right),
\end{equation}
and which approaches ${\rm Var}[\partial_\lambda w_\lambda]$ as $\lambda \to 0$.
As a given model is trained, so that the ${\rm ESS}$ approaches $1$, the result above shows that $\mathcal{E}^2 \to 0$ and an improvement in the variance of the estimated observable is expected, compared to 
the reweighting factor $w_\lambda^{\rm Id}$ in \Cref{eq:fdiff1} which corresponds to the identity flow transformation, $f = \mathrm{Id}$.
Under the independence assumption of \Cref{eq:vardO}, the ratio of $\mathcal E^2$ between the trained and identity flow can be used to predict the variance reduction for any measured derivative observables as
\begin{equation}
    \frac{\mathrm{Var}[d\left<\mathcal{O}\right>_{\rm Id}/d\lambda]}{\mathrm{Var}[d\left<\mathcal{O}\right>/d\lambda]} \simeq \frac{\mathcal E^2_{\rm Id}}{\mathcal E^2}.
    \label{eq:varredeq}
\end{equation}
The Feynman--Hellmann application involves reweighted estimates of two-point functions as inputs to the spectral decomposition, thus an analogous reduction in noise is expected from using flowed estimates of these two-point functions.

It is also possible to infer the expected volume scaling of the variance reduction. To see this, we note that the exponential volume scaling of the ESS~\cite{Abbott:2022zsh} at sufficiently large volumes implies, for small $\lambda$,
\begin{equation}
    {\rm ESS}(V) = 1 - k V \lambda^2 + O(\lambda^3),
\end{equation}
where $k$ is a constant. It then follows that
\begin{equation}
    \mathcal E^2 = k V + O(\lambda),
\end{equation}
so that $\mathcal E^2$ is extensive in the volume, and similarly for $\mathcal E^2_{\rm Id}$. Since the variance reduction can be approximated by the ratio appearing in \Cref{eq:varredeq}, and both $\mathcal E^2$ and $\mathcal E^2_{\rm Id}$ exhibit the same volume scaling, the variance reduction is expected to be approximately independent of the volume.

\subsection{Determinant ratio estimation}
\label{sec:PFs}

Several of the applications demonstrated in this work are evaluated for dynamical QCD.
While the inserted operators used in these applications are purely gluonic, a ratio of fermion determinants must be estimated to determine the reweighting factors in \Cref{eq:rw-lambda}. In this work, only the case of $N_f = 2$ degenerate light quark flavors are considered, in which case the $\lambda = 0$ action can be  written in terms of the lattice gauge action $S_g$ and the lattice Dirac operator $D$ as
\begin{equation}
\begin{aligned}
    S_0(U) = S_g(U) - \log \det D D^\dagger(U).
\end{aligned}
\end{equation}
The relevant reweighting factors are then
\begin{equation}
\begin{aligned}
    \log \hat{w}_{\lambda}(V) &= - S_{\lambda}(V) + S_0(U) - \log J(V) \\
    &= \lambda {\insertedOp}(V) - S_g(V) + S_g(U) \\
    &\qquad + \log \det\frac{DD^\dagger(V)}{DD^\dagger(U)} - \log J(V).
\end{aligned}
\end{equation}

To compute the fermionic contribution to the reweighting factor, the pseudofermion trick~\cite{Bhanot:1983uy} can be used to write the reweighting factor as an expectation value under Gaussian samples $\phi$,
\begin{equation}
    \det \frac{D D^\dagger(V)}{D D^\dagger(U)} = \left< e^{-\phi^\dagger (M M^\dagger)^{-1} \phi} \, e^{\phi^\dagger \phi} \right>_{\phi},
\end{equation}
where $M = [D(U)]^{-1} D(V)$ and $\left< \cdot \right>_\phi$ indicates an average with respect to the Gaussian distribution $\exp(-\phi^\dagger \phi)$.
This corresponds to the methodology of using ``naive pseudofermions'', which can lead to high variance in the reweighting factors. There are several ways to improve on this, as described in Ref.~\cite{Abbott:2022zhs}: one can use even/odd preconditioning~\cite{1988CoPhC..52..161D,Luscher:2010ae}, include more pseudofermion hits and average the reweighting factors, or use a pseudofermion flow.

In this work, pseudofermion flows are used. For this, the gauge-equivariant ``Parallel Transport Convolutional Network'' (PTCN) architecture discussed in Ref.~\cite{Abbott:2022zhs} is applied. In this architecture, a Gaussian pseudofermion field is transformed iteratively by taking linear combinations with its parallel-transported neighbors. Schematically,
\begin{align}
    \begin{split}
         \phi'(x) =& A \phi(x) + B U_{\mu}(x) \phi(x+\hat \mu)   \\&+ C U^\dagger_{\mu}(x-\hat \mu) \phi(x-\hat \mu) \  + ... ,
    \end{split}
\end{align}
where $A,B,C$ represent matrices of trainable parameters. By dividing the pseudofermion field into active and frozen degrees of freedom, a tractable Jacobian $J(\phi')$ can be computed. Furthermore, the PTCN architecture can easily be combined with even-odd preconditioning; this approach is employed throughout this work.
Given a trained pseudofermion flow, an improved estimate for the determinant ratio can then be obtained by evaluating
\begin{equation} \label{eq:pf-det-ratio}
    \det \frac{D D^\dagger(V)}{D D^\dagger(U)} = \left< e^{-\phi'^\dagger (M M^\dagger)^{-1} \phi'} \, e^{\phi^\dagger \phi} J(\phi')^{-1} \right>_{\phi}.
\end{equation}

\subsection{Linearization of the models} \label{sec:lin}

Although the value of $\lambda$  can be made small to control the $O(\lambda)$ bias in the derivative (\Cref{eq:dOdLflow} and \Cref{eq:dMpi0}), a more elegant and robust approach is to eliminate the bias entirely. This can be achieved through an exact linearization of the approach. An additional advantage is that the gauge fields are not changed and quark propagators do not need to be recomputed for each operator insertion, as will be discussed below.  

In the limit $\lambda \to 0$, the optimal flow $f(U)$ approaches the identity transformation plus an $O(\lambda)$ linear term. The behavior of the flow in this limit can be expressed in terms of the instantaneous \emph{flow field} $F(U)$ as
\begin{equation}
    V = \{ 1 + \lambda F(U)\} \, U + O(\lambda^2) = e^{\lambda F(U)} U + O(\lambda^2).
\end{equation}
Above, the flow field $F(U)$ is a field of anti-Hermitian, traceless elements of the group algebra describing the tangent direction of the flow expanded around the original gauge configuration $U$.

In this limit, it is therefore possible to exactly evaluate the derivative in $\lambda$ by expanding to first order. Expanding the weight first gives
\begin{equation} \label{eq:linear-what}
\begin{aligned}
    &\hat{w}_\lambda(V) = \frac{e^{-S_0(V) + \lambda {\insertedOp}}}{e^{-S_0(U)} \, J(V)} \\
    &= 1 + \lambda \, \big[ \nabla \cdot F - F \cdot \nabla S_0 + {\insertedOp} \big] + O(\lambda^2),
\end{aligned}
\end{equation}
where all terms in the expansion are evaluated on the gauge configuration $U$.
The operator $\nabla$ here indicates a vector of operators $\nabla^a_{x,\mu}$, so that
\begin{equation}
    F \cdot \nabla \equiv \sum_{\mu,x,a} F_{x,\mu}^a\nabla_{x,\mu}^a\,,
    \quad
    \nabla \cdot F \equiv \sum_{\mu,x,a} (\nabla_{x,\mu}^aF_{x,\mu}^a)\,.
\end{equation}
The normalized weight introduces an extra factor that gives the vacuum subtraction,
\begin{equation} \label{eq:linear-w}
\begin{aligned}
    &w_\lambda(V) = \frac{Z_0}{Z_\lambda} \hat{w}_\lambda = ( 1 - \lambda \langle {\insertedOp} \rangle + O(\lambda^2) ) \, \hat{w}_\lambda \\
    &\quad = 1 + \lambda \, \underbrace{\big[ \nabla \cdot F - F \cdot \nabla S_0 + {\insertedOp} - \langle {\insertedOp} \rangle \big]}_{\partial_\lambda w_\lambda} + O(\lambda^2).
\end{aligned}
\end{equation}
The observable can likewise be expanded as
\begin{equation}
    \mathcal{O}(V) = \mathcal{O}(U) + \lambda \big[ F(U) \cdot \nabla \mathcal{O}(U)\big] + O(\lambda^2).
\end{equation}

Combining these terms, an unbiased estimator (i.e., with no finite-difference approximation) for the $\lambda$ derivative of an arbitrary observable is finally given by (cf.~\Cref{eq:dOdLflow})
\begin{equation} \label{eq:linear-deriv}
\begin{aligned}
    \frac{d\langle \mathcal{O} \rangle}{d\lambda}\Bigg|_{\lambda=0} &= \left< (\partial_\lambda w_\lambda) \mathcal{O} + F \cdot \nabla \mathcal{O} \right>_0 \\
    &= \left< K \mathcal{O} \right>_0 + \langle {\insertedOp} \mathcal{O} \rangle_0 - \langle {\insertedOp} \rangle_0 \langle \mathcal{O} \rangle_0,
\end{aligned}
\end{equation}
where the linear operator
\begin{equation}
    K = \nabla \cdot F - F \cdot \nabla S_0 + F \cdot \nabla
\end{equation}
encodes the effect of the flowed estimate of the derivative.
The correctness of \Cref{eq:linear-deriv} follows because $K$ acts as a total derivative when taken in an expectation value with any smooth test function,
\begin{equation}
    \braket{K \mathcal{O}}_0 = \frac{1}{\mathcal{Z}_0} \int \mathcal{D} U \,
    \nabla \cdot \left(F \, \mathcal{O} \, e^{-S_0(U)}\right)
    = 0.
\end{equation}

For $N$-point functions evaluated using the derivative trick, the variance improvement obtained by the flowed definition of the derivative can be viewed in this light as the introduction of a particularly motivated control variate $K \mathcal{O}$, which vanishes in expectation but which should anti-correlate with ${\insertedOp} \mathcal{O}$ to reduce fluctuations of the estimator~\cite{Bacchio:2023all}.
Optimizing the flow drives $w_\lambda \to 1$, so that the linear term of the weight must approach zero,
\begin{equation} \label{eq:linear-transport}
    \partial_\lambda w_\lambda = \nabla \cdot F - F \cdot \nabla S_0 + {\insertedOp} - \langle {\insertedOp} \rangle \approx 0.
\end{equation}
In this limit the dependence on ${\insertedOp}$ in \Cref{eq:linear-deriv} approximately disappears and the derivative evaluation reduces to an $(N-1)$-point function,
\begin{equation}
    \frac{d \langle \mathcal{O} \rangle}{d \lambda} \approx \langle F \cdot \nabla \mathcal{O} \rangle_0.
\end{equation}
Correspondingly, for weights approximately independent of the observable, the prefactor ${\rm Var}[\partial_\lambda w_\lambda]$ determining the noise of the correlation function is driven to zero. This perspective also motivates a future training approach based on a Physics-Inspired Neural Network (PINN) loss function that directly minimizes \Cref{eq:linear-transport}; optimizing in this way is shown in \Cref{app:training} to be equivalent, at leading order in $\lambda$, to minimizing the KL divergence for a short-distance flow.

Similarly, for Feynman--Hellmann applications, the exact derivative of the two-point function (cf.~\Cref{eq:Ch}) can be computed as
\begin{equation}
\begin{aligned}
    \frac{dC^h(t)}{d\lambda}\bigg\rvert_{\lambda=0}
    = \Big\langle &
        \left( K + {\insertedOp} - \langle{\insertedOp}\rangle_0 \right)\, O^h_t\,[O^h_0]^\dagger
      \Big\rangle_0\,.
    \label{eq:linearized-3pt}
    \end{aligned}
\end{equation}
The matrix element of interest (cf.~\Cref{eq:FH}) may then be obtained from the time-dependence of the ratio
\begin{equation}
   \frac{dC^h(t)/d\lambda\big|_{\lambda=0}}{C_0^h(t)}
    \xrightarrow{\;t \gg 1\;}
    \text{const}
    + t\;\frac{dE_h}{d\lambda}\bigg\rvert_{\lambda=0}\,,
    \label{eq:ratio-slope}
\end{equation}
whose slope at large $t$ yields $dE_h/d\lambda$.
This is the standard summed-insertion method: because ${\insertedOp}$
is inserted on every timeslice, the sum over intermediate times
produces the characteristic linear growth in $t$ that projects onto
the ground-state matrix element.

In practice, the flow field associated with this infinitesimal limit can be obtained from a trained finite flow by evaluating the full flow transformation $V = f(U)$ and approximating 
\begin{equation} \label{eq:exactP}
    F(U) \approx \tfrac{1}{\lambda} \mathcal{P}[V U^\dagger].
\end{equation}
However, in this work the residual-flow architecture used provides an efficient shortcut to define a flow field $F(U)$ from a trained flow.
In this case, expanding the composition of residual layers that defines $V$ in \Cref{eq:exactP} gives up to $O(\lambda)$ 
\begin{equation} \label{eq:resid-expansion}
    V = U + \lambda \sum_k F^k_x(U) + O(\lambda^2),
\end{equation}
in terms of the flow fields $F^k_x$ defined for each layer in \Cref{eq:reslayer}.
Then, \eqref{eq:exactP} leads to the approximation
\begin{equation} \label{eq:FsumFk}
    F(U) = \sum_k F_x^k(U).
\end{equation} 
In this case, the exact divergence $\nabla \cdot F(U)$ can likewise be evaluated using the known Jacobian of the residual flow transformation.

As discussed above,
we empirically find that individual flow fields $F_x^k(U)$ take $O(1)$ values after training, which justifies the expansion in \Cref{eq:resid-expansion}.
While in principle one could imagine other parametric $\lambda$-dependence appearing in the layers of the trained model,
in practice this appears not to be preferred by the inductive bias of the parameterization used here. 

Finally, as for finite flows, an estimator involving the Dirac operator is required to evaluate the term $F \cdot \nabla S_0 = F \cdot \nabla S_g - F \cdot \nabla \log \det D D^\dagger$ for dynamical QCD applications. The latter term can be evaluated using either naive or flowed pseudofermions similarly to the finite flow case,
\begin{equation}
\begin{aligned}
    &F \cdot \nabla \log \det D D^\dagger = \Tr[(D^\dagger)^{-1} (F \cdot \nabla D D^\dagger) D^{-1}] \\
    &\quad = \left< (D^{-1} \phi)^\dagger (F \cdot \nabla D D^\dagger) D^{-1} \phi \right>_\phi \\
    &\quad = \left< (D^{-1} \phi')^\dagger (F \cdot \nabla D D^\dagger) D^{-1} \phi' J(\phi')^{-1} \right>_\phi.
\end{aligned}
\end{equation}
This is equivalent to the $O(\lambda)$ expansion of \Cref{eq:pf-det-ratio} used for the case of a finite flow, so that the noise properties of this estimate are equivalent for small $\lambda$.

\section{Summary of the workflow} \label{sec:workflow}

Below, a concrete workflow that may be used to train and evaluate derivative-trick observables is detailed for both the finite-difference approximation and the unbiased linearized estimator. This approach is used for the numerical applications discussed in \Cref{sec:glueballs} and \Cref{sec:structure}.

\subsection{Workflow for flowed derivative observables}

\paragraph{Using a finite-difference approximation.}
To estimate a derivative observable using a flow trained for a finite value of $\lambda$, the following workflow is used.
\begin{enumerate}
    \item Train a gauge-field flow $f$ at small lattice volume. In dynamical QCD, the small volume allows an exact evaluation of the fermion determinant to be used.
    \item For dynamical QCD: Train a pseudofermion flow at small volume.
    \item Apply flow $V_i = f(U_i)$ for $N$ gauge configurations $U_i$ at the target volume. Evaluate the corresponding weight $\hat{w}_\lambda$, either exactly (for pure-gauge YM) or using the flow-improved pseudofermion estimator of the determinant ratio (for QCD).
    \item Evaluate the derivative observable as
    \begin{equation}
        \frac{1}{N} \sum_i \left[ \frac{\hat{w}_\lambda(V_i)}{\tfrac{1}{N} \sum_j \hat{w}_\lambda(V_j)} \mathcal{O}(V_i) - \mathcal{O}(U_i) \right].
    \end{equation}
\end{enumerate}

\paragraph{Using linearization.}
To estimate a derivative observable using a linearized flow, providing an unbiased estimate of the $\lambda \to 0$ limit, the following workflow is used.
\begin{enumerate}
    \item Train a gauge-field flow $f$ at small volume. Extract the linearized flow field $F$.
    \item For dynamical QCD: Train a pseudofermion flow at small volume.
    \item Evaluate flow field $F_i = F(U_i)$ for $N$ gauge configurations $U_i$ at target volume. Evaluate the corresponding weight derivative
    \begin{equation}
        \partial_\lambda \hat{w}_\lambda = \nabla \cdot F_i - F_i \cdot \nabla S_0(U_i) + {\insertedOp}(U_i),
    \end{equation}
    either exactly (for pure-gauge YM) or using the pseudofermion estimator defined above (for dynamical QCD).
    \item Evaluate the unbiased derivative observable as
\begin{equation}
\begin{aligned}
    &\frac{1}{N} \sum_i \left[ (\partial_\lambda \hat{w}_{\lambda,i}) \mathcal{O}(U_i) + F_i \cdot \nabla \mathcal{O}(U_i) \right] \\
    &\quad - (\tfrac{1}{N} \sum_i \partial_\lambda \hat{w}_{\lambda,i} )(\tfrac{1}{N} \sum_{j} \mathcal{O}(U_j)).
\end{aligned}
\end{equation}
\end{enumerate}

\subsection{Cost comparison between linearization and finite differences}

In addition to the obvious advantage of eliminating $O(\lambda)$ effects in the linearized approach, there are other practical considerations. Notably, the finite-difference approach requires applying the full flow and evaluating observables and reweighting factors on the flowed gauge fields. In contrast, the linearized method only requires evaluating the observables and weights using the original gauge-field ensemble $\{ U_1, U_2, \dots \}$. This can be important when working with fermionic observables because Dirac operator inversions only need to be computed on the original, unflowed gauge configurations.

Another issue to consider is the total number of inversions required in each workflow. As an explicit example, consider the case of evaluating the meson three-point functions
\begin{equation}
    C^{(3)}_i(x) = \left< {\insertedOp}_i (\bar{\psi}_x \Gamma \psi_x) (\bar{\psi}_0 \Gamma' \psi_0) \right>
\end{equation}
for $n$ choices of operators ${\insertedOp}_1, \dots, {\insertedOp}_n$, where $\Gamma$ and $\Gamma'$ represent arbitrary Dirac gamma structures.

Following the finite-difference method, each three-point function can be obtained by applying the derivative trick to the two-point observable $\mathcal{O} = \left<(\bar{\psi}_x \Gamma \psi_x) (\bar{\psi}_0 \Gamma' \psi_0) \right>_\psi$ with distinct flows $f_1, \dots, f_n$ that have been respectively trained to target each inserted operator. Evaluating these $n$ correlation functions with finite flows requires evaluating
\begin{equation}
    \mathcal{O}(V) = \Tr\left[ D^{-1}_{0x}(V) \Gamma D^{-1}_{x0}(V) \Gamma' \right]
\end{equation}
by constructing the propagator\footnote{The transpose is obtained by $\gamma^5$-hermiticity.} $D^{-1}_{x0}(V)$ on the flowed fields $V$ for each flow, as well as for the unflowed fields---a total of $n+1$ propagators. 

In contrast, evaluating the same correlation functions by linearizing the flows requires calculating the term
\begin{equation}
\begin{aligned}
    &F \cdot \nabla \mathcal{O} = \\
    &\quad- \sum_{z,z'} \Big\{ \Tr\left[ D^{-1}_{0z} (F \cdot \nabla D_{zz'}) D^{-1}_{z'x} \Gamma D^{-1}_{x0} \Gamma'
    \right] \\
    &\qquad \qquad + \Tr\left[ D^{-1}_{0x} \Gamma D^{-1}_{xz'} (F \cdot \nabla D_{z'z}) D^{-1}_{z0} \Gamma'
    \right] \Big\},
\end{aligned}
\end{equation}
which can be computed following two different methodologies. First, one can use sequential inversion through the term $F \cdot \nabla D$. In this case, evaluating for each flow field $F \in \{F_1, \dots, F_n\}$ requires $n+1$ propagator inversions to access all sink times, i.e., the counting is identical to the finite-difference case. Second, one can use sequential inversion through the sink. In this case, the flowed correlation function can be computed for an arbitrary number of target operators by contracting with the respective flow fields $F_i \cdot \nabla D$, requiring $n_{\rm sink}+1$ propagator inversions, where $n_{\rm sink}$ is the number of sink times. For a large number of operators, as is typically the case for
insertions involving quark bilinears with different momenta, the latter approach is advantageous even if several sink times are used. These two approaches precisely echo the standard choice of inverting through the operator or sink to compute fermionic three-point functions.

This highlights the benefit of the linearized approach in this context: three-point functions can be evaluated using the same propagator inversions that would be used for the standard evaluation of three-point functions involving quark bilinear operators, while maintaining the variance-reduction provided by the flowed approach. This significantly reduces the inversions needed when a large number of operator choices are considered and allows reuse of any propagators that would already be calculated on the target ensemble for analogous three-point function calculations.
Despite this, for cases in which only a few operator insertions are considered, or for practical convenience in avoiding sequential inversions, a finite-difference calculation with controlled bias may be preferable to the linearized flow estimates. As such, in the remainder of this work both finite and linearized flow estimates are investigated.

\section{Numerical applications to glueball spectroscopy}\label{sec:glueballs}

This section presents applications of the methodology developed here to reduce the variance of glueball correlators. Specifically, correlators of the glueball with quantum numbers $J^{PC}=0^{++}$ (i.e., the scalar glueball) are studied in both SU(3) Yang-Mills theory and QCD in four space-time dimensions.

Using the derivative trick in \Cref{eq:derivative_trick}, the scalar glueball two-point function is connected to the derivative of a one-point function,
\begin{equation}
\begin{aligned}
    C(t) &= \langle \mathcal{O}(t) {\insertedOp}  \rangle_0 - \langle   \mathcal{O}(t)  \rangle_0  \langle   {\insertedOp}   \rangle_0 = \frac{d \langle \mathcal{O}(t) \rangle}{d \lambda} \Bigg \rvert_{\lambda=0},
    \label{eq:gluecorr}
\end{aligned}
\end{equation}
with $S_\lambda = S_0 - \lambda {\insertedOp}$ as in \Cref{sec:corrflows}. The interpolating operators used in the above definition are chosen to be
\begin{align}
    \mathcal O(t) &= \sum_{\vec x} \sum_{i\neq j=1}^3 \mathrm{Re} \Tr P_{ij}(\vec x, t),
     \quad {\insertedOp} = \mathcal O(0) \label{eq:interpglue},
\end{align}
with $P_{ij}(x)$ the (untraced) $1\times 1$ Wilson loop along the spatial $i,j$ directions.
The operator insertion has been localized on a particular timeslice, which ensures that the spectral decomposition of the resulting two point functions holds. To take advantage of time-translation averaging, the derivative-trick evaluations of $C(t)$ are repeated for all possible temporal translations of each gauge configuration. The effect this has on the overall computational cost is discussed in \Cref{sec:advan}.

The derivative definition of the vacuum-subtracted correlation function holds both in the pure gauge theory ($S_0 = S_{\rm YM}$) and in QCD ($S_0 = S_{\rm QCD}$).
Below, both applications are shown. A summary of models used is presented in \Cref{tab:modelcont}.

\begin{table*}
    \begin{ruledtabular}
    \begin{tabular}{ccccccccc}
    Model & Theory & Parameters  & Train geom.  & Train ESS / $\mathcal E^2$ & Eval.~geom. & Eval. ESS / $\mathcal E^2$  \\ \hline \noalign{\vskip 2pt}
    
    A &  SU(3) YM & $\beta=6.0$, {$\lambda=2\cdot 10^{-3} $ }& $4^3 \times 32$ & 99.996\% / $10$ & $16^3 \times 32$ & 99.6\% / {950} \\
    B
      & \multirow{2}{*}{$N_f=2$ QCD}
      & \multirow{2}{*}{$\beta=5.6, \kappa=0.1577, {\lambda=2\cdot10^{-3}}$}
      & \multirow{2}{*}{$4^3 \times 32$}
      & \multirow{2}{*}{99.996\% / $10$}
      & \multirow{2}{*}{$16^3 \times 32$}
      & 98.5\% / {3800} \\ 
   B+PF  & & & & & & 98.8\% / {3000} \\ 
    \end{tabular}
    \end{ruledtabular}
    \caption{Details of the trained models used for the demonstration of glueball correlator computation. For Model B, the two rows of evaluation metrics
    correspond to evaluations with naive pseudofermions (``B'') and with a pseudofermion flow model (``B+PF'').}
    \label{tab:modelcont}
\end{table*}

\subsection{Glueballs in Yang-Mills Theory}\label{sec:glue1}

The scalar glueball is first investigated in the SU(3) Yang-Mills theory. 
A flow model is constructed and trained as described in \Cref{sec:corrflows}. The architecture is defined as two stacks\footnote{A `stack' of layers is a set of layers such that every link is actively transformed once, composed.} of residual layers, the first with masking pattern ``mod 2'' and the second ``mod 4'', as described in Ref.~\cite{Abbott:2024kfc}. The target distribution is defined by fixing $\lambda=10^{-3}$. All models were trained using a $4^3 \times 32$ lattice geometry, then transferred to the target lattice volumes.
Model details and resulting quality metrics after training are reported as ``Model A" in \Cref{tab:modelcont}. The model was trained for approximately 10k optimizer steps, corresponding, in a non-optimized implementation, to about 3000 NVIDIA A100 GPU$\cdot$hours.

\begin{figure}
    \includegraphics[width=\linewidth]{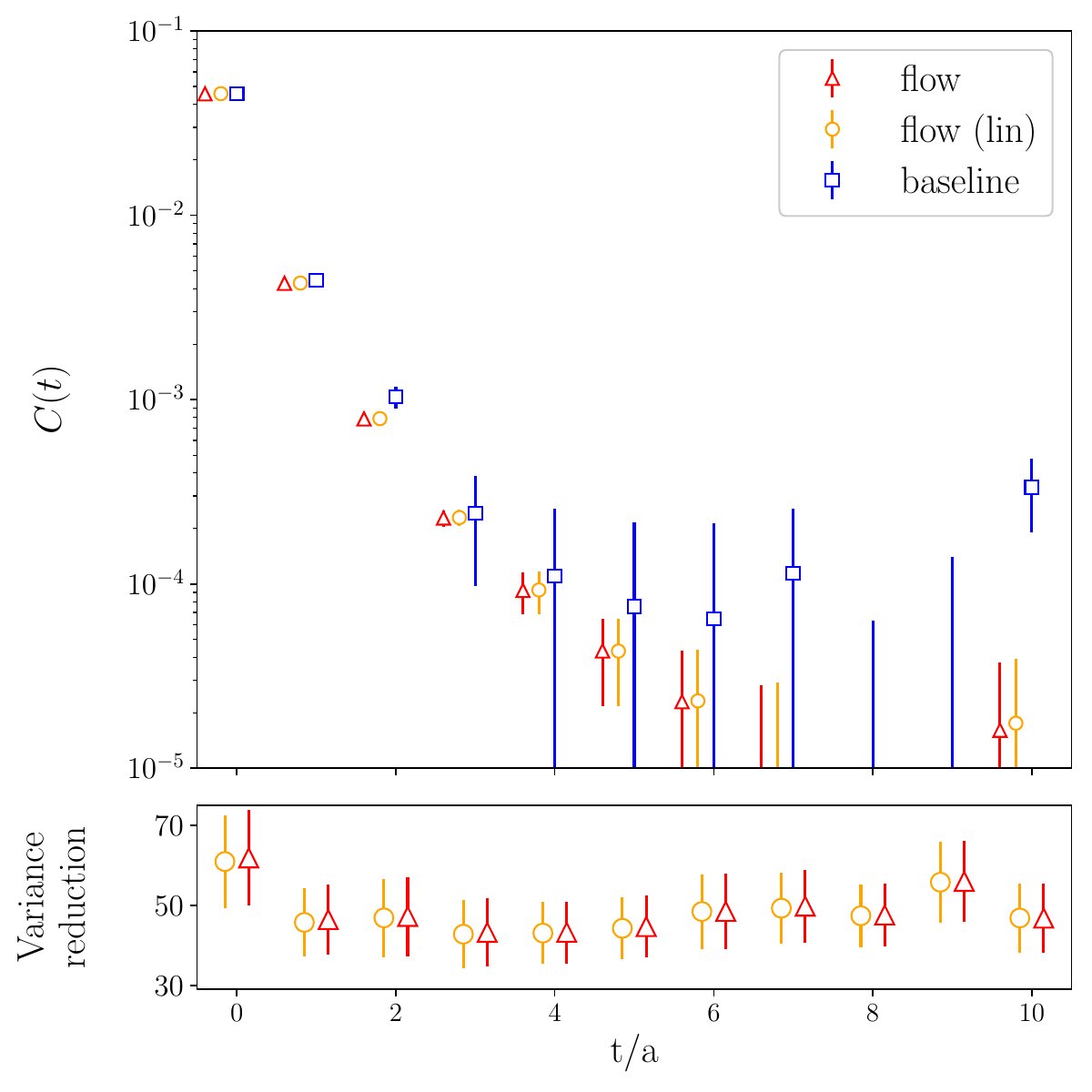}
    \caption{Illustration of variance reduction in the glueball correlator in the SU(3) Yang-Mills theory with $\beta=6$ for a $16^3 \times 32$ lattice geometry. The upper panel displays the correlator itself, evaluated on an ensemble of 3200 configurations.
    The baseline calculation using the standard two-point function evaluation is compared against the derivative-trick estimate evaluated using both the finite flow with $\lambda = {2 \cdot 10^{-3}}$ and the unbiased linearized estimate. 
    In the lower panel, the ratio between the variance of the baseline and flow-improved estimates is shown.
    }
    \label{fig:glueball1}
\end{figure}

The results for the correlation function evaluated at the largest target lattice geometry of $16^3 \times 32$ are shown in \Cref{fig:glueball1}. Here, both the finite-difference and linearized strategies outlined in \Cref{sec:workflow} are compared against a baseline result obtained using the standard translation-averaged, vacuum-subtracted evaluation of the two-point function in \Cref{eq:gluecorr}. The correlation functions computed using the flow model with both approaches are consistent with the baseline correlation functions within uncertainties for all timeslices. 
However, the flow-improved estimates give a constant variance reduction of a factor of around 50 across all timeslices, resulting in a measurable signal persisting for two additional timeslices.
The results for the finite-$\lambda$ case are also indistinguishable from the linearized case in terms of both central value and variance reduction. This nontrivial check verifies the equivalence of the two approaches in the limit $\lambda \to 0$.

The ratio $\mathcal E_{\rm Id}^2 / \mathcal{E}^2$ should provide the leading-order approximation to the observable-specific variance reduction, under the independence approximation in \Cref{eq:vardO}. 
For this example at the evaluation volume, the identity flow has $\mathrm{ESS}=89\%$, which leads to $\mathcal E_{\rm Id}^2 = 3.1 \cdot 10^4$. The ratio is then $\mathcal E_{\rm Id}^2 / \mathcal{E}^2 \simeq 32$, which is consistent with the order of magnitude of the observed variance reduction.

Another metric demonstrating the efficacy of the flow approach is the precision of the resulting effective mass estimates. As a representative example, the effective mass defined using the correlator at timeslices 1 and 2 is calculated,
\begin{equation}
    a m_{\rm eff} = - \log [C(2a)/C(a)], 
\end{equation}
giving the flow-improved and baseline values
\begin{align}
\begin{split}
     a m_{\rm eff}({\rm flow}) &= 1.695(23), \\
     a m_{\rm eff}({\rm baseline}) &= 1.48(15). 
\end{split}
\end{align}
Both results are consistent within $2\sigma$, and the variance in the flow case is approximately a factor of 40 smaller.
While this observable is still affected by excited state effects, and therefore does not directly estimate the glueball mass, it is used to emphasize that the observed variance reduction is not isolated to the correlator and persists in derived quantities.

Next, the hypothesis that the reweighting factors drive the variance of the correlation function can be tested by comparing the variance of the flowed estimates with and without the inclusion of the corresponding reweighting factors. For the glueball correlator studied here, the unweighted correlation function is found to be significantly more precise than the reweighted correlation function, albeit with a growing bias particularly at large values of $t$. 
This explains the behavior of the observed flow-improved correlation function at the present level of training: the flow successfully models the short distance correlation functions, but mismodels the tail, requiring increasing contributions from noisy reweighting factors to provide the correct expectation value. This highlights an opportunity for further gains, as a more expressive flow with further training could reduce the deviation from the exact correlation function, requiring significantly smaller and thus more precise reweighting factors.

\begin{figure}
    \includegraphics[width=\linewidth]{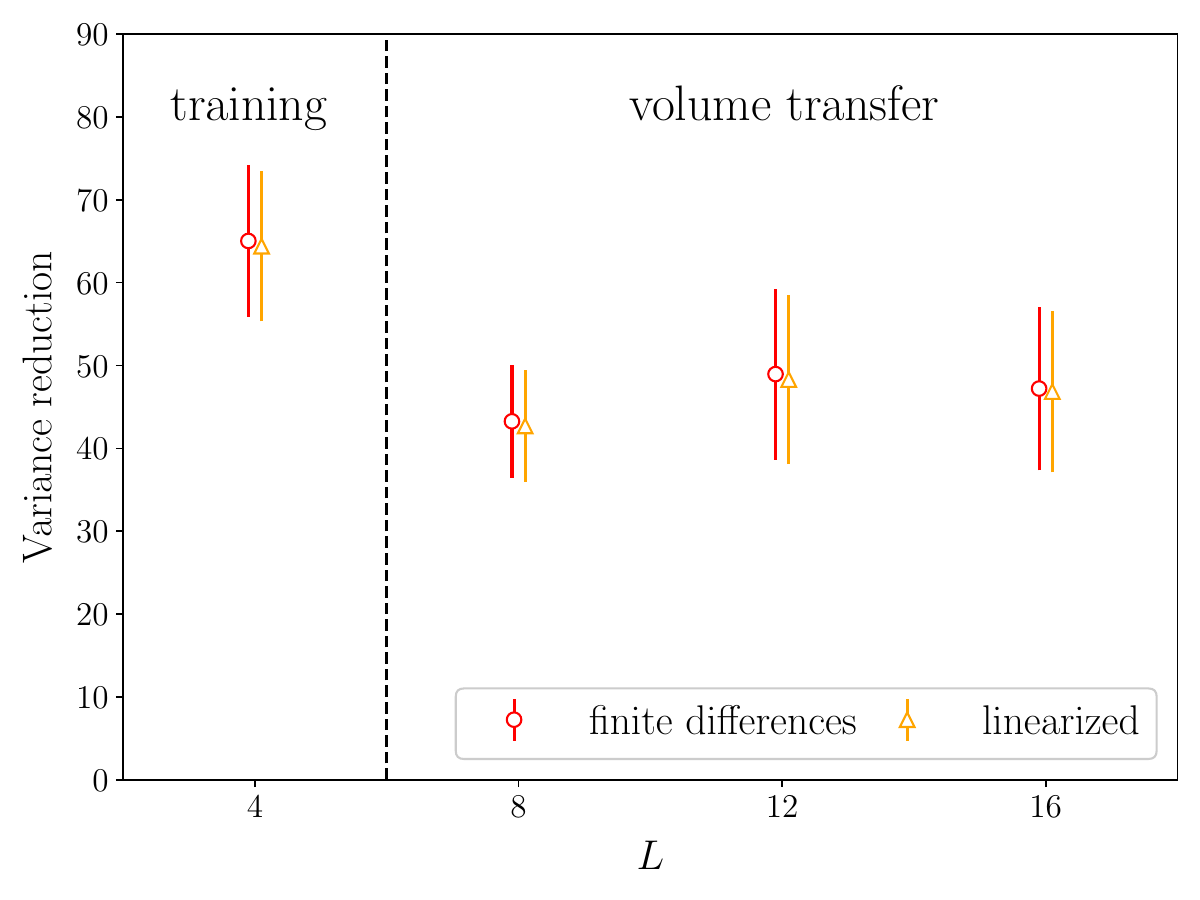}
    \caption{Variance reduction in the glueball correlator achieved by the flow approach at timeslice $t=2a$ as a function of the $L^3 \times 32$ lattice geometry. Results for both the finite-$\lambda$ and linearized approach are shown. Data for the $L=16$ case is obtained from an ensemble of 3200 configurations while the remaining cases are evaluated using ensembles of approximately 1600 configurations each. }
    \label{fig:varredballs}
\end{figure}

Finally, the volume-dependence of the flow-improved correlator is investigated by evaluating the trained flow across additional lattice geometries. While it is well-known that normalizing flows used for direct sampling deteriorate exponentially with increasing volume~\cite{Abbott:2022zsh}, in \Cref{subsec:derivviaflow} we have argued why this is  not expected for the variance reduction application. Indeed, the numerical results here also confirm this.
In \Cref{fig:varredballs}, the variance reduction for a fixed timeslice is shown for calculations using lattice geometries $L^3 \times 32$ ranging from the training geometry with $L=4$ to the largest target geometry with $L=16$. The net gain achieved by a fixed flow---measured by the variance reduction factor---is approximately constant with volume. The results are nearly identical for both the finite-$\lambda$ and linearized approaches. This demonstrates the viability of training at small volumes, then transferring to evaluate the resulting models at significantly larger volumes.
This strategy minimizes training costs with minimal loss in performance.

\subsection{Glueballs in QCD}\label{sec:glue2}

The scalar glueball correlation function
is next investigated in QCD. In this case, the same interpolating operators defined in \Cref{eq:interpglue} are used to define the correlation function, but the underlying theory is taken to be QCD with $N_f=2$ Wilson fermions. The action parameters are set to $\beta=5.6$ and ${\kappa=0.1577}$, corresponding to $aM_\pi=0.2392(11) $, ${M_\pi L \simeq 3.8}$, ${a \simeq 0.077}$ fm and $M_\pi \simeq 600$ MeV. 

In this setting, training the flow model requires significantly more computation per training step
than in the Yang-Mills example, because inversions of the Dirac operator are needed to generate training samples with Hybrid Monte Carlo.
This is mitigated first by working again with a small lattice volume for the training step. The cost is further controlled by exploiting model transfer, which reduces the total number of training iterations until convergence. In particular, the trained model from \Cref{sec:glue1} is used as the initial model, from which only 1k additional training steps are applied with the dynamical QCD target action to reach convergence. To reduce the noise in the gradients, this training is performed using the explicit determinant of the Dirac operator. While this is feasible in terms of memory use for the training volume ($4^3 \times 32$), it also incurs additional computational costs. Overall, with these choices, the computational cost per gradient step during training is approximately ten times larger than in the pure-gauge case, while the number of training steps is ten times smaller, leading to a roughly equivalent training cost.

In contrast to the pure gauge case, the computation of reweighting factors in the evaluation volume involves a pseudofermion estimate of the determinant ratio, as explained in \Cref{sec:PFs}. This typically leads to a lower ESS in the evaluation volume than an exact determinant-ratio estimator. Here two strategies are compared: a naive pseudofermion estimator with even/odd preconditioning, and a determinant-ratio estimator based on a pseudofermion flow model.  The latter leads to a higher ESS with a fixed number of pseudofermion hits with a negligible overhead in evaluation costs, bringing the final ESS closer to the value that would be obtained by exactly evaluating the determinant ratio.
The resulting model details and quality metrics after training are reported as ``Model B'' (using naive pseudofermions) and ``Model B+PF'' (using a pseudofermion flow model) in \Cref{tab:modelcont}.

\begin{figure}
    \includegraphics[width=\linewidth]{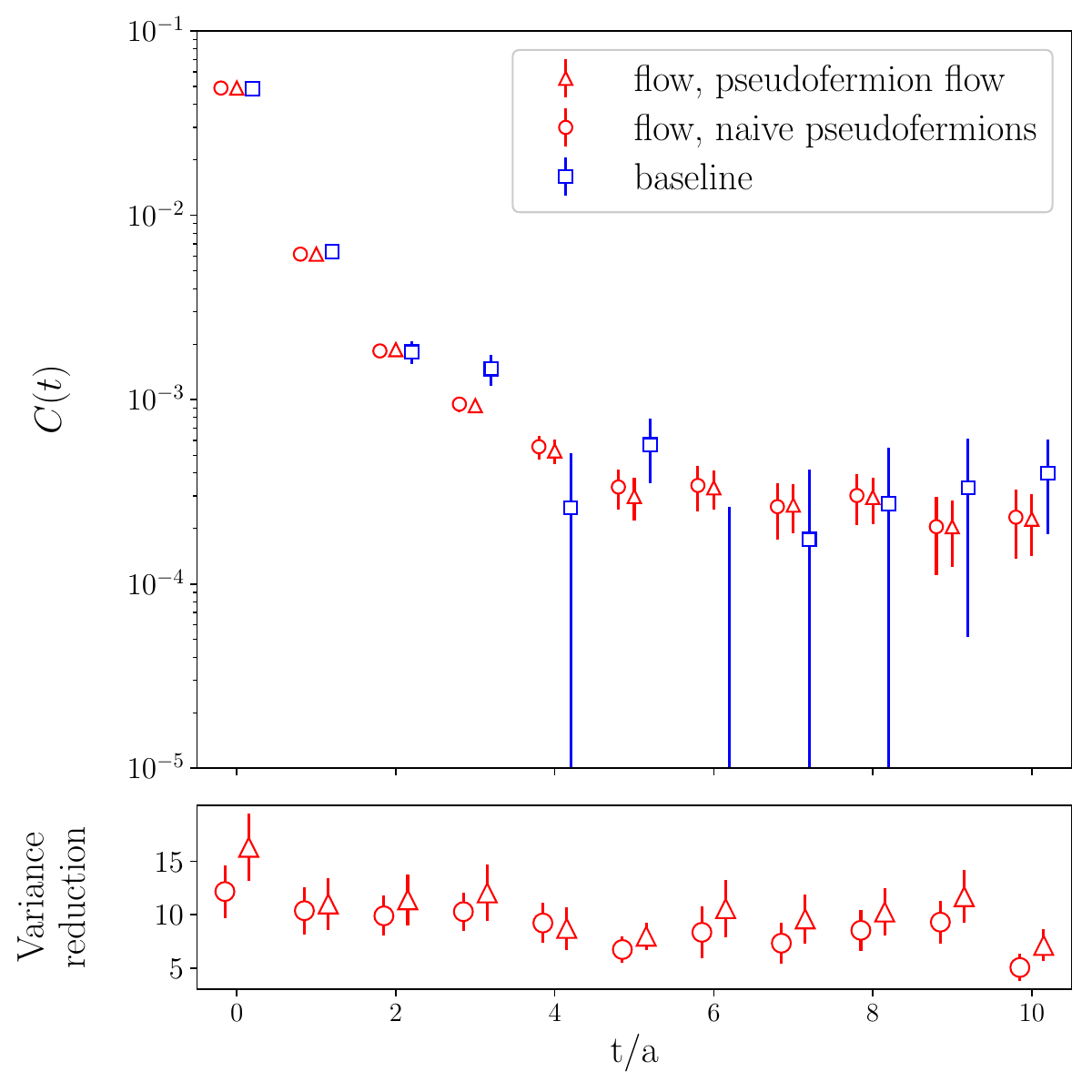}
    \caption{
    Illustration of the variance reduction in the glueball correlator in QCD with $N_f=2$ Wilson fermions using 1280 configurations. The upper panel displays the correlator itself;
    the flow-improved estimates using both naive pseudofermions and a pseudofermion flow are compared against a baseline calculation using the standard computation of the two-point function.
    A single pseudofermion hit for both flow estimates.
    In the lower panel, the ratio of variances is shown with and without employing a pseudofermion flow. 
    }
    \label{fig:glueball2}
\end{figure}

The results are shown in \Cref{fig:glueball2}, both with naive pseudofermions and with a pseudofermion flow model, using a single pseudofermion sample per gauge configuration in both cases. Even with naive pseudofermions, a significant variance reduction by a factor of 5--10 is obtained at the level of the correlation function, extending the signal region by one to two timeslices. Including a pseudofermion flow model further reduces the variance by roughly 20\%.

Computing the effective mass between $t=a$ and $t=2 a$, as in the Yang-Mills case, also demonstrates an improvement in precision. The effective mass estimates for the flow estimate with naive pseudofermions, the flow estimate with a pseudofermion flow model, and the baseline estimate are
\begin{align}
    \begin{split}
        a m_{\rm eff}({\rm flow}) &= 1.212(39), \\  
        a m_{\rm eff}({\rm flow, PF}) &= 1.193(33), \\
        a m_{\rm eff}({\rm baseline}) &= 1.26(15).
    \end{split}
\end{align}
Relative to the baseline, the variance is reduced by a factor of 14 using with naive pseudofermions, and 20 with a pseudofermion flow model.

\section{Numerical application to hadron structure}\label{sec:structure}

Another class of derivative operators of great phenomenological interest are hadronic matrix elements, which can be expressed in the form of derivatives through the Feynman--Hellmann theorem as in \Cref{eq:FH}. This procedure assumes that the operator is inserted in every timeslice, for which time-translation invariance is preserved.

Here, this approach is applied to a study the gluon momentum fraction of the pion in dynamical QCD.
A gluonic operator is used for this demonstration as this kind of operator suffers particularly severely from signal-to-noise challenges, and variance reduction on a finite ensemble may thus be particularly important in this case. The approach can also be extended to quark operators.

\begin{table*}
    \begin{ruledtabular}
    \begin{tabular}{ccccccccc}
    Model & Theory & Parameters  & Train geom.  & Train ESS / $\mathcal E^2$ & Eval. geom. & Eval. ESS / $\mathcal E^2$ \\ \hline \noalign{\vskip 2pt}
    C & \multirow{2}{*}{ QCD }
      & \multirow{2}{*}{$\kappa = 0.164111, \beta=3.8, \mu=0.012, \lambda=2\times 10^{-3}$}
      & \multirow{2}{*}{$4^4$}
      & \multirow{2}{*}{99.94\% / 150}
      & \multirow{2}{*}{ $16^3 \times 32 $} 
      &  65.0\% / $1.3\times 10^5$ \\
    C+PF  & & & & & & 69.2\% / $1.1\times 10^5$ \\ 
    \end{tabular}
    \end{ruledtabular}
    \caption{Details of the trained models used for the demonstration of hadronic matrix element calculation via the Feynman--Hellmann theorem.
    The two rows of evaluation metrics
    correspond to evaluations with naive pseudofermions (``C'') and with a pseudofermion flow model (``C+PF'').
    }
    \label{tab:modelcont2}
\end{table*}

The gluon momentum fraction of the pion is computed following the methodology of Ref.~\cite{QCDSF:2012mkm}. Specifically, the operator considered is a linear combination of components of the discretized Energy-Momentum Tensor (EMT),
\begin{equation}
    {\insertedOp}_{\rm gEMT} = -\frac{\beta}{N_c} \sum_x \text{Tr Re }\left( \sum_{i} P_{i0}  - \sum_{i<j}  P_{ij} \right) \ ,
\end{equation}
where $i,j \in \{1,2,3\}$.
The matrix element of this operator, obtained in the pion rest frame via \Cref{eq:FH}, is related to the gluon momentum fraction of the pion $\langle x \rangle_g$ by
\begin{equation}\label{eq:mastereqFH}
  \frac{d M_\pi}{d\lambda}   \bigg \rvert_{\lambda \to 0} = \frac{1}{2M_\pi} \langle \pi \vert {\insertedOp}_{g\text{EMT}} \vert \pi \rangle =-\frac{3M_\pi}{2} \langle x \rangle_g^{\rm latt} \ ,
\end{equation}
where ``$\text{latt}$'' indicates that it is a bare matrix element.
The full action $S_\lambda = S_0 - \lambda {\insertedOp}_{\rm gEMT}$ can be expressed as an anisotropic action, where temporal and spatial plaquettes have different couplings. This is an amenable target for the residual layers, as shown in Ref.~\cite{Abbott:2024kfc}. 

Following this methodology, the gluon momentum fraction is determined in $N_f=2$ QCD using twisted-mass fermions and the tree-level improved gauge action. This setup, chosen for simplicity and to reduce discretization effects, is described in Ref.~\cite{Urbach:2007rt}, where the lattice spacing is determined to be approximately $a=0.10$ fm. The parameters are
$\kappa = 0.164111$, $\beta=3.8$, and $\mu=0.012$, 
yielding a pion mass of approximately $M_\pi \simeq 540 $ MeV. The target distribution is defined by fixing $\lambda=2 \times 10^{-3}$.

All models were trained using a $4^4$ lattice geometry, then transferred to a target lattice geometry of $16^3 \times 32$. Model details and resulting quality metrics after training are reported as ``Model C'' (using naive pseudofermions) and ``Model C+PF'' (using a pseudofermion flow model) in \Cref{tab:modelcont2}; in both cases three pseudofermion samples are used to evaluate reweighting factors.  The model was trained for approximately 6k
optimizer steps, corresponding, in a non-optimized implementation, to about 4k NVIDIA A100 GPU·hours.

The standard two-point correlation function is constructed on the flowed and un-flowed ensembles via \Cref{eq:Ch}. Fits in the range $[t_{\rm min}, T/2]$ yield the pion mass, the mass derivative is estimated from the finite difference between the results at $\lambda = 2 \times 10^{-3}$ and $\lambda = 0$, and the gluon momentum fraction is then extracted by \Cref{eq:mastereqFH}.

\begin{figure}
    \includegraphics[width=\linewidth]{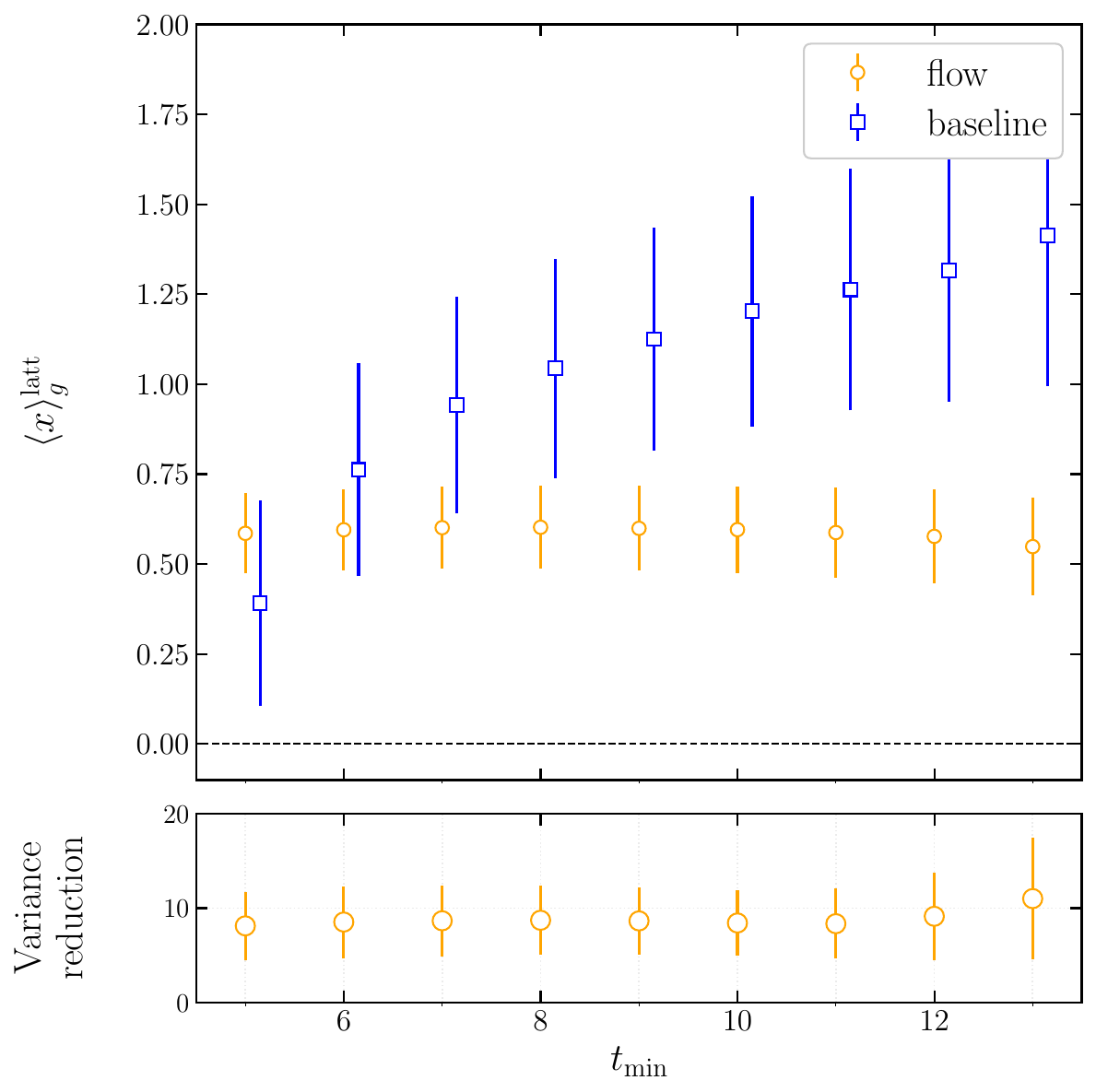}
    \caption{Illustration of the variance reduction in the gluon momentum fraction in QCD with $N_f=2$ Wilson fermions. The upper panel compares the gluon momentum fraction of the pion computed using flows (orange squares) versus the baseline $\epsilon$-reweighting method (blue circles), both evaluated on the same ensemble of 8576 configurations.
    In the lower panel, the ratio between the variance of the baseline and the flow-improved estimate is shown.
    The results are shown as a function of the lower end of the fit range used for the extraction of the mass, $t_{\rm min}$.}
    \label{fig:EMT}
\end{figure}

The results are shown in \Cref{fig:EMT}, where the gluon momentum fraction as a function of $t_{\rm min}$ is shown.
To provide a baseline against which to measure variance improvement, results are compared to those computed by taking the derivative of the mass directly~\cite{Bouchard:2016heu} (implemented in practice via $\epsilon$-reweighting~\cite{Abbott:2024kfc}).
The variance of the momentum fraction estimate is reduced relative to the baseline by a factor of approximately 10 by using the flow approach. Moreover, the cost of the flow approach (using finite differences) is only slightly higher: while correlation functions need to be evaluated using the flowed fields, the inversions from each unflowed field can be used to initialize the solver on the corresponding flowed field to accelerate this calculation. As such, the cost of the flowed approach is less than a factor of 2 larger than the standard approach.

\section{Discussion of computational advantage}\label{sec:advan}

The results of \Cref{sec:glueballs,sec:structure} have demonstrated significant variance reduction in several observables computed on a fixed number of gauge fields, achieved by combining the derivative trick with application of flow models. However, using the flow methodology also increases the computational cost. Thus, assessing whether the demonstrated variance reduction represents true computational advantage requires considering all aspects of the computational pipeline, including the training and application of a flow model, configuration generation, measurement, and reweighting. Moreover, beyond sheer computational costs, there are other factors to consider: storage needs for configurations and propagators, as well as data management of large numbers of measurements (which is particularly relevant for e.g., studies of glueballs which may use a very large number of gauge fields).

A general feature of the examples considered in this work is that training costs are marginal: as a result of the small lattice volume used for training, these costs are smaller than generation of the target-volume configurations and for these examples are negligible in the computational budget. Moreover, even using a small training volume, training costs can be mitigated by transfer learning, as shown in the QCD glueball example,
and simultaneous calculations of multiple observables can exploit this to amortize training costs even further. Rather than the training, it is thus the \textit{application and reweighting} associated with the flow model that are potentially significant costs.

For the first example considered---glueballs in Yang-Mills theory---the cost of generating configurations via heatbath/overrelaxation and the cost of performing measurements and reweighting are low, as there are no fermions and so no inversions are necessary. The dominant cost in this case is applying the flow transformation, and therefore the flow approach is significantly more expensive than standard methods. This is in part due to the fact that the operator is inserted in each timeslice, to take advantage of time-translation averaging, so the flow must be applied $T$ times. While the variance reduction by a factor of roughly $50$ is significant, this falls short of compensating for the additional computational overhead for the implementation used here by approximately a factor of $5$. Presumably, additional investment in flow architectures and training could improve upon this to the vicinity of the break-even point.

Here, however, the primary advantage of using flows can be manifest in another aspect: calculations of glueballs in pure gauge theories tend to use very large ensembles, e.g.\ $10^7$ gauge configurations in Ref.~\cite{Abbott:2025irb}. As such, a reduction of a factor of 50 in the number of configurations required can be valuable, even if computational costs break even. Fewer configurations lead to reduced storage costs, less data movement, and therefore lighter analysis pipelines, which can be of great significance at these data volumes.

This is precisely the advantage obtained in the calculation of glueballs in QCD. While the variance reduction achieved by using flows is in this case only a factor of $10$, configuration generation costs are more significant. With the current flow model quality, the net result is approximately equal cost of flowed and unflowed approaches at fixed target error, with an order of magnitude fewer configurations required in the flow case. This demonstrates already a practical advantage, and further optimization of the flow models can yield both practical and computational gains. 

Finally, the case of the hadron structure application in this work demonstrates clear computational advantage with the current flow model quality. Here, the flow evaluation and reweighting costs are subdominant for two primary reasons: the flow must only be applied once per configuration because of time-translation invariance and propagators are computed on several sources per configuration to evaluate correlation functions. 
In this context, the cost of the finite-difference version of the flow pipeline is only marginally higher than the standard pipeline, because the inversions on unflowed fields can be reused to initialize solves on the corresponding flowed fields; the additional inversion only adds a 30--50\% overhead in cost. A similar cost is expected for the linearized evaluation, for which the inversions required are analogous to standard calculation of three-point functions.
For a pipeline where the cost of HMC is dominant, reducing the variance by a factor of 10 directly yields a 10$\times$ computational advantage. For a measurement-dominated pipeline, the advantage is approximately a factor of 7. For the particular implementations used in this demonstration, the advantage is approximately a factor of 8.

Taken in sum, the examples demonstrate that current flow technology can yield both computational and practical advantages in lattice QCD calculations. The precise magnitude of computational efficiency gain at a fixed target precision depends intimately on the details of the calculation and model used---as well as the efficiency of its implementation---but for the present implementation and applications, factors of a few have been demonstrated. In all cases, however, the flow approach yields an important practical advantage in that it allows for a variance reduction given a fixed-size ensemble of gauge fields. As ensembles are stored and re-used as community resources, a fixed-size ensemble is in many cases typical, and improved variance in this setting represents a useful objective. This advantage is magnified towards the continuum limit and physical quark masses, where costs of ensemble generation are more significant.

\section{Conclusion and outlook}\label{sec:concl}

This work demonstrates the applicability of machine-learned flows as a tool to reduce the variance of lattice QCD calculations.
This approach is based on formulating observables of interest as derivatives with respect to action parameters---a method that is applicable to arbitrary $N$-point functions using the generating functional approach. These calculations can then be interpreted as a reweighting problem between nearby distributions, where machine-learned flows are effective at reducing the variance of the reweighting factors by a learned change of variables. In this work, variance reductions by factors of 10--100 are demonstrated for a range of applications involving inserted gluonic operators, including glueball correlation functions and matrix elements in the context of hadron structure. An unbiased estimator for flowed derivative observables is also constructed by linearizing the machine-learned flow, removing a potential source of systematic error in the method. The generality of the approach, applied here for the first time for dynamical QCD quantities in four-dimensional lattice geometries up to $16^3 \times 32$, suggests that the flow-based variance reduction procedure is potentially useful for a wide range of applications in lattice QCD.

The computational advantage of this methodology has also been analyzed for each example studied. In the case of hadron structure, a lower uncertainty is obtained with the flow approach for the same number of measurements, resulting in a computational advantage of a factor of approximately 8. In the glueball case, the flow approach is comparable or slightly more expensive, but with clear practical advantages and the potential of additional computational gains with further refinement. 

An important conclusion of this work is that the advantage achieved by utilizing flows for variance reduction does not deteriorate when transferring to larger volumes. This was explicitly illustrated for the case of glueballs in Yang-Mills theory in \Cref{fig:varredballs}, and it is expected to hold for other cases. As such, training can be performed at small volumes at very low computational cost that is negligible for the examples investigated here.

While this work studies dynamical QCD, all considered operator insertions are gluonic. An important future extension is therefore to explore fermionic operators, such as scalar or vector currents. This extension is conceptually straightforward but will require some new flow architecture development for the operator insertion. Similar success in this setting would yield variance reduction in observables such as sigma terms or the hadronic vacuum polarization tensor.

Ultimately, this work has demonstrated computational and practical advantages of a flow approach in the evaluation of various phenomenologically-interesting observables in lattice QCD. This opens a new avenue of at-scale applications, in which machine-learned flows can now be included in the standard lattice QCD toolkit.

\vspace{1cm}

\section*{Acknowledgements}

We thank Michael Albergo, Kyle Cranmer, Mathis Gerdes, Alberto Ramos, and Ross Young for useful discussions.

RA is an Ernest Kempton Adams Postdoctoral Fellow supported in part by the Ernest Kempton Adams fund for Physical Research of Columbia University and in part by U.S. DOE grant No. DE-SC0011941.
RA, DB, YF, PES and JMU are supported in part by the U.S.\ Department of Energy, Office of Science, Office of Nuclear Physics, under grant Contract Number DE-SC0011090. PES is additionally supported by the U.S.\ DOE Early Career Award DE-SC0021006, by a NEC research award, and by the Carl G and Shirley Sontheimer Research Fund. The work of FRL was supported in part by the Platform for Advanced Scientific Computing (PASC) project ``ALPENGLUE''. 
This document was prepared using the resources of the Fermi National Accelerator Laboratory (Fermilab), a U.S. Department of Energy, Office of Science, Office of High Energy Physics HEP User Facility. Fermilab is managed by Fermi Forward Discovery Group, LLC, acting under Contract No. 89243024CSC000002.
This work is supported by the U.S.\ National Science Foundation under Cooperative Agreement PHY-2019786 (The NSF AI Institute for Artificial Intelligence and Fundamental Interactions, \url{http://iaifi.org/}). It is also associated with an ALCF Aurora Early Science Program project, and has used resources of the Argonne Leadership Computing Facility which is a DOE Office of Science User Facility supported under Contract DEAC02-06CH11357.

The authors acknowledge the MIT SuperCloud and Lincoln Laboratory Supercomputing Center~\cite{reuther2018interactive} for providing HPC resources that have contributed to the research results reported within this paper. Part of the calculations were performed on UBELIX (\url{https://www.id.unibe.ch/hpc}), the HPC cluster at the University of Bern. 

Numerical experiments and data analysis used PyTorch~\cite{paszke2019pytorch}, JAX~\cite{jax2018github}, Haiku~\cite{haiku2020github}, Horovod~\cite{sergeev2018horovod}, NumPy~\cite{harris2020array}, and SciPy~\cite{2020SciPy-NMeth}. Figures were produced using matplotlib~\cite{Hunter:2007}.

\appendix
\crefalias{section}{appendix}

\section{Loss functions and optimization} \label{app:training}
All models in this work are trained using the Adam optimizer applied to path gradient estimates of the reverse Kullback-Leibler divergence between the model density $q(V) = r(U) \times J(V)$ and the target density $p(V) \propto e^{-S_\lambda(V)}$.\footnote{{Here we also assume that in the case of a dynamical theory, the action $S_\lambda$ includes the effects of the fermions, either directly through the log fermion determinant or through a pseudofermion action. Thus, all results in this appendix also hold for QCD.}}
Explicitly, the loss function and gradient estimates using the path gradient approach are defined in terms of expectation values with respect to $r(U) \propto e^{-S_0(U)}$ as
\begin{align} \label{eq:loss-pg}
    &\mathcal{L}^{\rm KL} = \left< \log J(V) + S_\lambda(V) { - S_0(f^{-1}(V))} \right>_0, \nonumber \\
    &\nabla_\theta \mathcal{L}^{\rm KL} = \\
    & \quad \big\langle (\nabla_\theta V) \cdot \nabla_V (\log J(V) + S_\lambda(V) {- S_0(f^{-1}(V)))})
    \big\rangle_0, \nonumber
\end{align}
where {$\nabla_{\theta} V \equiv P[\frac{\partial f(U)}{\partial \theta} f(U)^\dagger]$ indicates the algebra-valued} gradient of $V = f(U)$ with respect to the parameters $\theta$ that define the flow $f$ and $\nabla_V$ indicates the algebra-valued left Lie gradient with respect to the variable $V$. The dot product is here taken to be between the components in the algebra of $\nabla_\theta V$ and $\nabla_V (\dots)$.

While the above loss function is used for all numerical applications in this work, a second loss function has also been investigated for directly training flow fields $F(U)$. The observation that $\partial_\lambda w_\lambda \approx 0$ should hold for an optimized flow (cf.~\Cref{eq:linear-transport}) suggests that a flow field could alternatively be trained by directly minimizing this quantity. This leads to a Physics-Inspired Neural Network (PINN) loss function---one that aims to solve a partial differential equation by direct optimization---which is expressed as
\begin{equation}
\begin{aligned}
    \mathcal{L}^{\rm PINN} &= \left< (\partial_\lambda w_\lambda)^2 \right>_0 \\
    &= \left< \left( \nabla \cdot F - F \cdot \nabla S_0 + {\insertedOp} - \langle {\insertedOp} \rangle \right)^2 \right>_0.
\end{aligned}
\end{equation}
All quantities in the expectation value are evaluated for the fields $U$ sampled from $r(U)$.
An analogous loss function has previously been proposed for generative modeling via annealing~\cite{Albergo:2024trn} and for directly optimizing control variates of lattice field theory observables~\cite{Bacchio:2023all}.

Remarkably, this loss function is closely related to the KL divergence. Simplifying the gradient of the PINN loss,
\begin{equation}
\begin{aligned}
    \nabla_\theta \mathcal{L}^{\rm PINN} &= 2 \left< \partial_\lambda w_\lambda \nabla_\theta \left( \nabla \cdot F - F \cdot \nabla S \right) \right>_0 \\
    &= -2 \left< \nabla_\theta F \cdot (\nabla \partial_\lambda w_\lambda) \right>_0,
\end{aligned}
\end{equation}
where the second line is obtained by integrating by parts with respect to the gradient $\nabla$. Expanding the gradient in \Cref{eq:loss-pg} with respect to $\lambda$, one finds that the gradients are related according to
\begin{equation}
    \nabla_\theta \mathcal{L}^{\rm PINN} = \lim_{\lambda \to 0} \frac{2}{\lambda^2} \nabla_\theta \mathcal{L}^{\rm KL},
\end{equation}
i.e., up to a rescaling by a factor of $\lambda^2 / 2$, the loss functions are equivalent to leading non-zero order. A slightly more involved derivation also shows that the loss functions themselves satisfy the same relationship. This suggests a future workflow in which \emph{flow fields} $F(U)$ are directly constructed and trained, without training and linearizing a finite flow. While an extensive exploration is left for future work, initial tests indicate that the approach works similarly to the reverse Kullback-Leibler training.

\bibliographystyle{utphys}
\bibliography{main}

\end{document}